\documentstyle[prb,epsf,floats,aps,amsfonts,twocolumn,eqsecnum]{revtex}

\begin{document}

\twocolumn[\hsize\textwidth\columnwidth\hsize\csname
@twocolumnfalse\endcsname

\title{ A Free Field Representation of the $Osp(2|2)$ current algebra at
level $k=-2$, \\
and  Dirac Fermions in a random $SU(2)$ gauge potential}

\author{Andreas W.W. Ludwig }
\address{
\cite{permaddress}
Department of Physics,  
University of California, Santa Barbara, CA 93106, and, \\
Institute for Theoretical Physics, Valckenierstraat 65, 1018 XE Amsterdam (The Netherlands)
}

\date{\today}

\maketitle

\begin{abstract}
The $Osp(2|2)$ current algebra at level $k=-2$ is known to describe  the
IR fixed point of 2D Dirac fermions, subject to  a random $SU(2)$ gauge
potential. We show that this theory has a simple  free-field representation
in terms of a compact, and a non-compact free scalar field, as well
as  a  free fermionic ghost, at $c=-2$. The fermionic twist fields are crucial 
for the construction. The logarithmic current-algebra
primary field  with vanishing scaling dimension,   
transforming in an indecomposable
representation,  appears as a consequence of familiar logarithmic operators
at $c=-2$.

\end{abstract}
\vspace{1cm}


]

\bigskip

\bigskip
\bigskip

\section{Introduction}

The current algebra based on the two-dimensional
$Osp(2|2)$ Wess-Zumino-Witten model at level $k=-2$,
is known\cite{Ludwigunpub,Bernard}
to provide a supersymmetric description of the
the infrared fixed point,  existing in the problem
of two species of Dirac Fermions
in two spatial dimensions, subject to 
a  random, static and short-ranged  $SU(2)$-gauge potential.
The latter problem has been studied extensively by various authors,
including the work by  Mudry, Chamon, and Wen\cite{MCW}, and 
by    Caux, Taniguchi and Tsvelik\cite{Tsvelik}.
A discussion of the $Osp(2|2)$ current algebra at general level $k$ was given
by Maassarani and Serban\cite{MS}. (The central charge is  $c=0$.) 
Here we show that at level $k=-2$,
relevant for the random gauge model, special simplifications occur:
at this level the currrent algebra and its primary fields 
 can be represented  in terms of a compact and a non-compact free
scalar field (each at central charge $c=1$), as well as a free fermionic
ghost at central charge $c=-2$. In order to obtain the representations
of the $Osp(2|2)_{-2}$ current algebra, it is crucial to include the
twist fields of the fermionic ghost  sector.
As pointed out by Gurarie\cite{Gurarie93},
 this $c=-2$ sector\cite{logs}
 contains  a  zero-dimensional
logarithmic operator, reflecting  a 
Jordan-block structure of the Virasoro generators.
In our free field representation it is  this  same
operator, which gives rise to the zero conformal weight 
$Osp(2|2)_{-2}$ current algebra primary field
transforming in an indecomposable representation.

\section{Motivation} 
\label{tmatr}
 The conformal weights of
 the $Osp(2|2)_{-2}$ current algebra, 
corresponding to Kac-Moody primary
fields transforming under $Osp(2|2)$
  in the (`typical')  representation
labeled by $[{\bf b},  {\bf q}]$
are given by\cite{MS,SNR}
$$
\Delta = { ( {\bf q}^2 - {\bf b}^2)/2},
$$
where ${\bf q}=0, 1/2, 1,.. $ and  we consider ${\bf b}$ 
real.  
We wish to exhibit a free field representation of 
$Osp(2|2)_{-2}$ in  terms of (i): a   compact free boson\footnote{
Conventional  normalization: 
$<\varphi(z,{\bar z}) )\varphi(0)>=-\ln (z {\bar z})$.}
$\varphi$, obtained by bosonization of 
the $SU(2)_1$ sub-current algebra,
(ii): a non-compact boson\footnote{
We choose $ i{\varphi'}$ real with conventional correlator:
${<i {\varphi'}(z,{\bar z}) \ i {\varphi'}(0)>=}-\ln (z {\bar z})$.}
$\varphi'$,  and (iii): a system of free fermionic
ghosts\cite{FMS,Gurarie93,Kausch} $\chi\equiv \chi_-$,
 $\chi^\dagger\equiv \chi_+$ at central charge $c=-2$,
normalized such that
$$
<\chi_a (z, {\bar z}) \chi_b  (w, {\bar w}) >=
\epsilon_{ab} \ln (z-w) ({\bar z}-{\bar w})
$$
($\epsilon_{ab}=$ $-\epsilon^{ab}=$ antisymmetric,
$\epsilon_{-  + }=1$.)
[From now on we will consider only the left-moving (holomorphic)
sectors, denoting $\varphi_L(z)$ by $\varphi$, ${\varphi'}_L(z)$
by ${\varphi'}(z)$ etc..]
The total central charge
$c=1+1-2$ adds up to zero.
In the $c=-2$ sector it will be crucial to also consider the fermionic twist fields
 $\mu$ and $\sigma_{a}$, $a= \pm$,  of
conformal weights $-1/8$ and $+3/8$, respectively.
To motivate the possibility of describing $Osp(2|2)_{-2}$
within the tensor product of the three free theories above,
consider the 4-dimensional representation
$[{\bf b}, {\bf q}]=[0,1/2]$ of conformal weight $\Delta=1/8$.
The representation decomposes under the bosonic subalgebra
into an $SU(2)$ doublet with charge  ${\bf b}=0$,
 and two $SU(2)$ singlets with  charge ${\bf b}=\pm 1/2$.
We claim that the doublet may be represented by
$$
v_{\alpha}\equiv
e^{ \alpha { 1\over \sqrt 2} i \varphi}  \ \  \mu,
\qquad
 ( \alpha = \pm ) 
$$
\begin{equation}
\label{fourdoublet}
({\rm conformal \  weights}: \ \ \Delta= 1/4 -1/8 = 1/8),  
\end{equation}
where $ \alpha/2 =\pm 1/2$ is the z-quantum number of $SU(2)$, 
and the singlets by
$$
w_{a} \equiv
e^{ a  { 1\over \sqrt 2} i{ \varphi'} }  \ \  \sigma_{a},
\qquad 
( a=\pm )
$$
\begin{equation}
\label{foursinglets}
({\rm conformal \  weights}: \ \  \Delta=-1/4 + 3/8 =1/8),
\end{equation}
where $a/2= \pm 1/2$ measures the charge quantum number ${\bf b}$.
Note that the conformal weights add up as required.

\section{Current algebra}  

In order to establish the free field representation,
we first show that the
eight  $Osp(2|2)_{-2}$
currents can be entirely expressed in terms of
the free fields.
The four bosonic currents consist of
three $SU(2)_1$ currents $Q_{++}, Q_{- -},
 Q_z$,
 together with the charge\footnote{This corresponds to the
non-compact sector described by $\varphi'$.}
current ${B}$, measuring the quantum number ${\bf b}$. These can be
written as
$$
Q_{++} =
  e^{ +{ 2\over \sqrt 2} i \varphi},
\qquad
Q_{- -} = 
e^{ - { 2\over \sqrt 2} i \varphi},
$$
\begin{equation}
\label{bosoniccurrents}
Q_{z} =  {1\over \sqrt{2} }
 \  i \partial \varphi,
\qquad
B=
{-1\over \sqrt{2} } i \partial {\varphi'}
\end{equation}
The fermionic currrents consist of two global $SU(2)$ doublets, one
($ V_{\alpha}$) 
with
charge ${\bf b}=+1/2$, and one ($ W_{\alpha}$) with
charge ${\bf b}=-1/2$.
We claim that these can be written as
$$
V_{\alpha}
=
e^{ + { 1\over \sqrt 2} i  ( \alpha \varphi  + { \varphi'}) }  
\ \partial \chi^\dagger
$$
\begin{equation}
\label{fermioniccurrents}
W_{\alpha}
 =
e^{ + { 1\over \sqrt 2} i  ( \alpha   \varphi  - { \varphi'}) }  \ 
\partial \chi
\end{equation}

It is easily checked that the bosonized expressions defined  above satisfy the
OPE's of the $osp(2|2)_{-2}$ current algebra (listed in the Appendix),
hence proving the claim.

\section{Four-dimensional Representation ($\Delta=1/8$)}

Next we establish that the
OPE's of the currents (as defined in the previous paragraph) 
 with the four 
fields defined in Eq.'s (\ref{fourdoublet},\ref{foursinglets}), 
are indeed those required for the corresponding Kac-Moody
primary field $[{\bf b}, {\bf q}]=[0,1/2]$.
The OPE's with the
bosonic currents (\ref{bosoniccurrents}) are obviously correct.
In considering the OPE's with the fermionic currents (\ref{fermioniccurrents}),
the twist fields of the $c=-2$ fermionic ghost sector 
are crucial. A useful tool in handling those
 is an extra global `isospin'  $SU_\chi(2)$ symmetry
of the $c=-2$ sector\cite{Kausch}, 
 whose generators we denote by $J_{++}, J_{- -},  J_3$,  under
 which the fermionic ghosts  transform
as a doublet:
$$
[J_3,  \partial \chi_\pm] = \pm {1\over 2} \partial \chi_\pm,
\quad
[J_{++} ,  \partial \chi_+] =
[J_{- -} ,  \partial \chi_-] = 0,
$$
$$
[J_{++},   \partial \chi_-] = \chi_+,
\quad
[J_{- -},   \partial \chi_+] = \chi_-
$$ 
 The OPE's  of the fermionic ghosts with the twist fields
are\cite{Kausch}:
\begin{equation}
\label{modesmu}
\partial \chi_\pm(z) \mu(w) \sim { (1/ \sqrt{2})  \over (z-w)^{1/2} } 
 \sigma_{\pm}(w)
\end{equation}
\begin{equation}
\label{modessigma}
\partial \chi_\pm (z) \sigma_{\mp}(w) 
\sim   {\mp  (1/\sqrt{2})  \over (z-w)^{3/2}}   \mu(w)
\end{equation}
when we adopt the following  normalizations of the 2pt. functions
\begin{equation}
\label{normsigmamu}
<\sigma_a(z) \sigma_b(0)>={ \epsilon_{ab}\over z^{3/4}},
\qquad
<\mu (z) \mu(0)>={1 \over z^{-1/4}}
\end{equation}
More generally ($n\geq 0$),
$$
\partial \chi_+(z) \sigma_{+n}(w)
 \sim   {\bar C}_n \   (z-w)^{n-1/2}  \  
 \sigma_{+(n+1)}(w),
$$
$$
\partial \chi_-(z) \sigma_{-n}(w) \sim 
C_n \   (z-w)^{n-1/2}  \  
  \sigma_{-(n+1)}(w)
$$
where
${\bar C}_n, C_n$ are constants.
Here
$\sigma_{\pm n}$ is the $c=-2$ Kac degenerate field 
of conformal weight $\Delta_{n+1, 2}= 2 ({n\over 2})^2 - {1\over 8}$,
which is a  highest weight
 under the global isospin  $SU_\chi(2)$ symmetry, 
 with quantum numbers $ j_3= \pm j =\pm n/2$.

Using Eq.'s(\ref{fermioniccurrents}, \ref{modesmu}, \ref{modessigma})
one easily verifies the OPE of the
eight $Osp(2|2)_{-2}$ currents,  collectively
denoted by ${\cal J}_{\cal A}$
( where ${\cal A}$ is the adjoint index),
with the
four-dimensional KM primary $
\phi_{\bf a}$:
\begin{equation}
\label{fourdimcurrrentope}
{\cal J}_{{\cal A}} (z) \phi_{\bf a}(0)
=
{1  \over z } \ 
{{(t_{\cal A})}^{\bf b }}_{\bf a }
 \ \ \phi_{\bf b}
   + ...
\end{equation}
 The representation matrices 
${{(t_{\cal A})}^{\bf b }}_{\bf a }$
for the generators  $V_{\pm}, W_{\pm}$
in the 4-dimensional representation (index ${\bf a}, {\bf b},...$)
 under consideration
 are e.g.  given in Ref.[\onlinecite{MS}] in the basis
$ (\phi_1, \phi_2, \phi_3, \phi_4)=$
$(v_+, w_-, w_+,  v_-)$.
 Since the Knizhnik-Zamolodchikov (KZ)
equation is a consequence of the OPE
(\ref{fourdimcurrrentope}), the four point functions
of the bosonized expressions given
in ( \ref{fourdoublet},\ref{foursinglets}) need to satisfy\footnote{
We also  see this explicitly below.}
the KZ equation given in Ref.[\onlinecite{MS}]. 

It is interesting to consider the OPE of two fields in the four-dimensional
representation.
Group-theoretically\cite{MS,GLR}, the tensor product (16 states) decomposes
into an 8-dimensional adjoint and another 8-dimensional, but 
indecomposable representation. 
This can be studied by looking at the
correlator of four fields $v_{\alpha}$ and $w_a$,
representing the  components of the conformal block of $ Osp(2|2)_{-2}$ primary
fields
in the representation $[{\bf b}, {\bf q}]= [0,1/2]$.
We now  focus on this correlator.

\subsection{Four-point function (conformal blocks), for $\Delta=1/8$}

Consider first the correlator of four $v_{\alpha}$, 
which may be decomposed into the two $SU(2)$ invariant 
tensors
$$
{( {\bf  I^2}  )}_{\alpha_1, \alpha_2, \alpha_3, \alpha_4}
=
\epsilon_{\alpha_1, \alpha_2} \ 
\epsilon_{\alpha_3, \alpha_4},
\ \ \ 
{ ( {\bf I^1})}_{\alpha_1, \alpha_2, \alpha_3, \alpha_4}
=
\epsilon_{\alpha_4, \alpha_1} \ 
\epsilon_{\alpha_2, \alpha_3}
$$
yielding\footnote{This decomposition follows from Eq.(\ref{fourdoublet}).}:
$$
< v_{\alpha_1}(z_1) v_{\alpha_2}(z_2) v_{\alpha_3}(z_3) v_{\alpha_4}(z_4)>=
$$
$$
=
[
{ ({\bf I^2} )} \  G_{\bf 2} \ 
+
{( {\bf I^1} )} \  G_{\bf 1} 
]
 \  <\mu \mu \mu \mu>
$$
Here 
$G_{\bf j}$ are correlators of four free boson exponentials.

Evaluation of $<v_- v_+ v_+ v_->$ and 
$<v_- v_- v_+ v_+>$ gives
$$
G_{\bf 2}
= 
{\bigl [ 
{1\over z_{13} z_{24} }
 \ 
{
 ( 1-\xi)
\over \xi
}  
\bigr ] }^{1/2},
\quad
G_{\bf 1}
= 
{\bigl [ 
{1\over z_{13} z_{24} }
  \ 
{ \xi \over (1-  \xi) }  
\bigr ] }^{1/2}
$$
where
$$
\xi= {z_{12} z_{34} \over z_{13} z_{24} },
\qquad
\bigl (  z_{ij}\equiv z_i-z_j \bigr )
$$
is a crossratio.
The two-dimensional space of  conformal blocks 
of the twist field $\mu$
is described in terms of hypergeometric functions\cite{Gurarie93}
$$
<\mu(z_1) \mu(z_2) \mu(z_3) \mu(z_4) >=
$$
$$
={[ z_{13} z_{24} \ 
\xi (1-\xi)]}^{1/4} f(\xi), 
$$
where $f(\xi)$ is an arbitrary linear combination of
the two functions
$$
 \ \ \ \ \ f^{(0)} (\xi) = F[1/2,1/2;1;\xi]= 1+ \xi/4+ ...  \qquad \qquad
$$
and
\begin{equation}
\label{twistconformalblock}
f^{(1)} (\xi)
=f^{(0)} (1-\xi) = f^{(0)} (\xi) \ \ln(\xi) + h(\xi)
\end{equation}
[$h(\xi)$ is a function,   regular at $\xi=0$].
Note that $f^{(1)}(\xi)$ has a logarithmic singularity at $\xi=0$.
The conformal block in question can therefore  be written in
the following simple  form
$$
< v_{\alpha_1}(z_1) v_{\alpha_2}(z_2) v_{\alpha_3}(z_3) v_{\alpha_4}(z_4)>=
$$
\begin{equation}
\label{fourblockdirect}
=
{ 1 \over [ z_{13} z_{24} ]^{1/4} }
\  
{\bigl [
{ 1 \over \xi (1-\xi)}
\bigr ] }^{1/4} \ 
   \bigl \{
 {\bf I^2} \  (1- \xi)   \ 
+
 \ {\bf I^1} \  \xi  
\bigr \}
\ f(\xi)
\end{equation}

\subsection{Knizhnik-Zamolodchikov equation of the current algebra,
for $\Delta=1/8$}

It is instructive to make connection with the  description of the
same conformal block in terms of current algebra\cite{MS}.
In general, the $Osp(2|2)_{-2}$ conformal block in the 4-dimensional representation
is a sum of three $Osp(2|2)$
invariant tensors\footnote{to be distinguished from 
the two $SU(2)$ tensors above (in boldface).}
 $I_1, I_2, I_3$,  each multiplied by a
coefficient function,  $F_1, F_2, F_3$, respectively.
These  latter functions satisfy the Knizhnik-Zamolodchikov (KZ) equation. 
Since these three functions appear for example in the following components,
it is easy to compare with the bosonized expressions above:
$$
<v_- v_- v_+ v_+>
=
<
\phi^{({ 1\over 8})}_{4}
\phi^{( {1\over 8})}_{4}
\phi^{( {1\over 8})}_{1}
\phi^{( {1\over 8})}_{1}
>=
$$
$$=
 [ {1 \over z_{13} z_{24} } ]^{1/4} \ [ {1\over 1 - \xi } ]^{1/4}
\ F_1(z),
$$
$$
<v_- v_+ v_+ v_->
=
<
\phi^{({1\over 8})}_{4}
\phi^{({1\over 8})}_{1}
\phi^{({1\over 8})}_{1}
\phi^{({1\over 8})}_{4}
>=
$$
$$=
 [ {1 \over z_{13} z_{24} } ]^{1/4} \ [ {1\over 1 - \xi}]^{1/4}
\ F_2(z), 
$$
and
$$<v_{-} v_{+} w_- w_+>=
<
\phi^{({ 1\over 8})}_{4}
\phi^{( {1\over 8})}_{1}
\phi^{( {1\over 8})}_{2}
\phi^{( {1\over 8})}_{3}
>=
$$
$$
=
 [ {1 \over z_{13} z_{24} } ]^{1/4} \ [ {1\over 1 - \xi } ]^{1/4}
\ (-1)  F_3(z), 
$$
where
$$
 - z={z_{12} z_{34} \over z_{23} z_{14}}=
{\xi \over 1 - \xi},
\qquad
\xi ={z \over z-1}
$$
is another  crossratio. 

Comparison with (\ref{fourblockdirect}) shows that
\begin{equation}
\label{constraint}
F_2(z) = {-1\over z} F_1(z)
\end{equation}
As discussed in the Appendix, this  condition
is consistent with the $Osp(2|2)$ Knizhnik-Zamolodchikov
(KZ) equation. The KZ-equation implies the following two additional
statements (see Appendix). First, when parametrizing
\begin{equation}
\label{parametrization}
F_1(z) = z^{3/4} (z-1)^{-1/4} \ f(z)
\end{equation}
$f(z)$ satisfies
the defining second order differential equation for the elliptic function
 $F[1/2,1/2;1;z]$.
Second, the KZ equation  also implies
\begin{equation}
\label{Fthreeg}
F_3(z)
=
({-4\over \alpha}) z^{3/4} (z-1)^{3/4} \
[ {1\over z} f(z) + 2 {d \over dz} f(z)]
\end{equation}
Using standard identities for elliptic functions\cite{Gradshteyn}
one finds that this expression is again related to
the solution of a hypergeometric equation\footnote{
$\alpha$ is here a parameter related to group-theory conventions,
and is  denoted by ${1\over \epsilon\gamma}$ in Ref.[\onlinecite{MS}]}:
$$
F_3(z)
=
({4\over \alpha})
z^{-1/4} (z-1)^{-1/4} \ g(z)
$$
 where
$$
\quad
g^{(0)}(z) =F[-1/2, 1/2;1;z],
 \ \  g^{(1)}(z) =
g^{(0)}(1-z)
$$
Note that the same expresssion for $F_3$ is obtained by using the bosonized
expressions in Eq.(\ref{fourdoublet},\ref{foursinglets}).
Hence one can see directly that the bosonized expressions satisfy the
relevant KZ equation, as expected from the OPE's.

\subsection{Single-valued Combinations}

Since the only non-trivial conformal blocks arise from the $c=-2$ sector,
the problem of combining holomorphic and antiholomorphic conformal blocks
is resolved by forming off-diagonal combinations of the two conformal blocks,
in the same way as at $c=-2$, in Ref.[\onlinecite{Gurarie93}].

\section{Eight-dimensional indecomposable representation and logarithms
($\Delta=0$)}

As mentioned above, the OPE of two 4-dimensional representations
$(\Delta=1/8)$, decomposes group theoretically into an adjoint
and an indecomposable representation. Since the adjoint channel
does not contain an $Osp(2|2)$ singlet, the singlet will
necessarily be a component of the 8-dimensional indecomposable,
of conformal weight $\Delta=0$. The latter results from the logarithmic
conformal block of the $c=-2$ twist field\cite{Gurarie93}
of Eq.(\ref{twistconformalblock}) above.
In order see this explicitly, consider the $SU(2)$-singlet
channel of (\ref{fourblockdirect}):
$$
{  \epsilon^{\alpha_1\alpha_2} \epsilon^{\alpha_3\alpha_4}
\over 4}
< v_{\alpha_1}(z_1) v_{\alpha_2}(z_2) v_{\alpha_3}(z_3) v_{\alpha_4}(z_4)>=
$$
$$=
{1\over [z_{12} z_{34}]^{1/4} } \ \{
C_0 \  f^{(0)}(\xi) 
+ C_1  \ f^{(1)}(\xi) 
+ O(\xi) \}=
$$
$$=
{1\over [z_{12} z_{34}]^{1/4} } \ \{
C_0 +
C_1 \ [\ln z_{12} + \ln z_{34} - 2 \ln z_{24}] + ...
\} 
$$
Hence the OPE in the singlet channel is
$$
{ \epsilon^{\alpha_1\alpha_2}  \over 2} \ 
v_{\alpha_1}(z_1)
v_{\alpha_2}(z_2)
\sim
$$
\begin{equation}
\label{fourfourindecomposableOPE}
\sim {1\over (z_{12})^{1/4}}
\{
[(\ln z_{12}) \  \phi_s(z_2) + \phi_t(z_2)] + ...\}
\end{equation}
with the two point functions
$$
<\phi_s(z_2) \phi_s(z_4)>=0,
$$
$$
<\phi_s(z_2) \phi_t(z_4)>=
<\phi_t(z_2) \phi_s(z_4)>= C_1, 
$$
\begin{equation}
\label{indecomposabletwoptfcts}
<\phi_t(z_2) \phi_t(z_4)>= - 2 C_1  \ \ln z_{24} + C_0
\end{equation}
Here $\phi_s, \phi_t$ are the `bottom-' and `top-'
component of the  indecomposable multiplet, respectively,
 (of zero charge $B$, and zero
spin) of Ref.[\onlinecite{MS}].
The constant $C_0$ arises from the freedom to redefine $\phi_t$
by addition of $\phi_s$ with an arbitrary coefficient. This
freedom gives rise to the presence of the conformal block $f^{(0)}$,
which is free of logarithms (in this channel).

Note that $\phi_t$ plays here the role of the identity operator,
since\cite{MS}
$$
<\phi_s>=0,
\qquad
<\phi_t>= {\rm const.}
$$
[The vanishing of the first expectation value  follows from supersymmetry --
similarly for the first correlator in Eq.(\ref{indecomposabletwoptfcts}).]
The fact that $\phi_t$ has an expectation value ensures
that the OPE of Eq.(\ref{fourfourindecomposableOPE})
implies
a `usual'
 non-vanishing two point function
$$
<v_{\alpha_1}(z_1) v_{\alpha_2}(z_2)>
\propto 
{ \epsilon_{\alpha_1\alpha_2}
\over
 (z_{12})^{1/4}}
$$

Similarly, in the spin-triplet channel (which lies in the
$Osp(2|2)$-adjoint)  the
current-algebra descendant of the indecomposable weight-zero
operator appears. The relevant OPE's can
be obtained in an analogous fashion, by considering
for example the $\xi\to 0$ limit of the correlator
$ <v_+ v_+ v_- v_->$ [from Eq.(\ref{fourblockdirect})].

\section{Other representations}

General representations of the $Osp(2|2)$ current
algebra  are labeled by $[{\bf b}, {\bf q}]$.
One expects that representations with 
$q\geq 1$ cannot occur at level $k=-2$,
because they would contain  inadmissible representations
of $SU(2)_1$  (that is,   primaries of this current algebra
with spin $q\geq 1$).

\section{Stress Tensor Multiplet} \label{supspin}

The total stress tensor of the $c=0$ theory is
the sum of the three free field stress tensors 
$$
T^{\varphi} = {1\over 4} (\partial\varphi)^2, \qquad
T^{\varphi'} = {1\over 4} (\partial{\varphi'})^2,
$$
$$
T^{\chi}=  {1\over 2} \epsilon^{ab} (\partial \chi_a)(\partial \chi_b)
= (\partial \chi^\dagger)(\partial \chi)
$$
$$
T=
T^{\varphi}+
(T^{\varphi'}+
T^{\chi}) 
\equiv T^f + T^b
$$
of central charge $c=+1, +1, -2$ respectively.
Clearly, the total stress tensor is that of the
$gl(1|1)$ current-subalgebra, generated by $W_+, V_-, Q_3, B$.

It was pointed out recently by Gurarie\cite{Gurarieb}
that in theories with (at least) a global $gl(1|1)$ symmetry
the stress-tensor transforms in the  (indecomposable)
adjoint  representation,
with the  `top'-component denoted by $t$. In the present case,
one finds
$$
t=
T^{\varphi}
-
(T^{\varphi'}+ T^{\chi})
\equiv  T^f - T^b
$$
Clearly, the so-defined stress tensors  $T^f$ and $ T^b$ commute, 
and have central charges $c=+1$ and $c=-1$, respectively.
The anomaly parameter of Ref.[\onlinecite{Gurarieb}] is therefore
$b=1$. The OPE of the stress tensor $T$ and its `companion' $t$
is not that of the logarithmic extension noted  recently
in the 2D percolation problem\cite{GurarieLudwig}.

\section{Conclusion}
 
We have presented a representation of the $Osp(2|2)$ current algebra at level
$k=-2$, in terms of two free scalar fields, one compact and the other non-compact,
as well as a free fermionic ghost sector at $c=-2$. We have shown how  
the 4-dimensional current algebra primary field (of conformal weight $\Delta =1/8$),
as well as the 8-dimensional primary field transforming in an indecomposable representation
(of conformal weight $\Delta =0$), appear in this representation.
All non-trivial structure arises from the twist fields of the $c=-2$ sector.
The four point function of the $\Delta=1/8$ field is shown to satisfy the
KZ equation.

\acknowledgements

I acknowledge discussions and/or  relevant previous collaborations
 with M. Jeng, I. Gruzberg,  S. Guruswamy, A. LeClair,  N. Read,
 K. Schoutens, and especially with V. Gurarie. 
I also acknowledge the hospitality of the
Institute for Theoretical Physics, at the University of Amsterdam,
during  a leave from Univ. of California, where this work was done.
This work was supported by the  NSF under  Grant DMR-00-75064.




\appendix

\section{OPE's of the $Osp(2|2)_{-2}$ current algebra}

The OPE's of the $osp(2|2)_{-2}$ current algebra
are\cite{SNR,MS}
$$
Q_{++} (z) Q_{- -}(0) \sim {1\over z^2} + {2\over z} Q_3
$$
$$
Q_3(z) Q_{++}(0) \sim {1 \over z} Q_{++}(0),
\quad
Q_3(z) Q_{- -}(0) \sim {-1 \over z} Q_{- -}(0),
$$
$$
Q_3(z) Q_3(0)  \sim {1/2\over z^2},
\qquad
B(z) B(0) \sim {-1/2\over z^2}
$$
$$
Q_{++}(z) V_-(0)\sim {1\over z} V_+(0),
\quad
Q_{++}(z) W_-(0)\sim {1\over z} W_+(0)
$$
$$
Q_{- -}(z) V_+(0)\sim {1\over z}  V_-(0),
\quad
Q_{- -}(z) W_+(0)\sim {1\over z}  W_-(0)
$$
$$
Q_3(z) V_{\pm}(0) \sim  {\pm 1/2 \over z} V_{\pm}(0),
\qquad
Q^3(z) W_{\pm}(0)  \sim {\pm 1/2 \over z} W_{\pm}(0)
$$
$$
B(z) V_{\pm}(0) \sim { 1/2 \over z} V_{\pm}(0),
\qquad
B(z) W_{\pm}(0) \sim {- 1/2 \over z} W_{\pm}(0)
$$
$$
W_+(z) V_+(0)\sim  {1\over z} Q_{++}(0),
\qquad
W_-(z) V_-(0) \sim {-1\over z} Q_{- -}(0)
$$
$$
V_+(z) W_-(0)
\sim
{-1\over z^2} + {B(0) - Q_3(0)\over z}
$$
$$
V_-(z) W_+(0)
\sim
{1\over z^2} + {-B(0) - Q_3(0)\over z}
$$

It may be useful to write
OPE's  involving $SU(2)$ tensor products
  in more compact form:
$$
Q_A(z) Q_B(0) \sim
{d_{AB}\over z^2} + {{f_{AB}}^C \over z} \ Q_C(0)
$$
$$
Q_A(z) V_{\alpha}(0) \sim { {{(\tau_A)}^\beta}_{\alpha}  \over z} \ V_{\beta}(0)
$$
$$
Q_A(z) W_{\alpha}(0) \sim { {{(\tau_A)}^\beta}_{\alpha}  \over z} \ W_{\beta}(0)
$$
$$
V_{\alpha}(z) W_{\beta}(0)
\sim
{\epsilon_{\alpha\beta} \over z^2}
+
{
\epsilon_{\alpha\beta} \  B(0)
+
{(\tau^A)}_{\alpha\beta} \ Q_A(0) \over z}
$$
Here, $A, B, C, ... \in \{ ++,  \ - -, \  3\}$ is  a spin-1 (adjoint)
index,  $d_{AB}$ and ${f_{AB}}^C$ are the metric and structure constants, respectively.
 $\alpha, \beta$ are spin-$1/2$ indices, 
and $ 
{
{(\tau_A)}^\alpha}_\beta$ is the representation matrix in 
the spin-1/2 representation (Pauli-matrices). Spinor indices are lowered and
raised with $\epsilon_{\alpha\beta}=-\epsilon^{\alpha\beta}$ and 
adjoint indices with $d_{AB}, d^{AB}$.

\section{Knizhnik-Zamolodchikov Equation}

In this Appendix we briefly discuss the
$Osp(2|2)$ Knizhnik-Zamolodchikov (KZ) equation(s)\cite{MS}
at level $k=-2$,
and their consequences. These equations are
$$
4 {d \over d z} F_1(z)
 = {- F_1 \over z(z-1)} +{ - {\alpha \over 2} F_3 \over z-1}, 
$$
$$
4 {d \over d z} F_2(z)
=
{1\over z (z-1)}
[ 2 F_1 + (-2z+3) F_2 + {\alpha \over 2 } F_3], 
$$
\begin{equation}
\label{KZappendix}
4 {d \over d z} ( {\alpha\over 2} F_3)
=
{ 1 \over z (z-1) }
[
2 ( - F_1 + (z-2) F_2) - {\alpha\over 2}  F_3], 
\end{equation}
where $\alpha$ is a parameter, denoted by ${1\over \epsilon \gamma}$
in Ref.[\onlinecite{MS}].

By adding the first two KZ equations one derives for 
$H(z)= F_1 + z F_2$ a simple equation  which,  upon integration, 
yields
\begin{equation}
\label{constraintappendix}
H(z) = {\rm Const.  } \ z^{1/4} (z-1)^{1/4}
\end{equation}
Furthermore, combining the first and the third KZ  equations,
and using (\ref{constraintappendix}) one finds
that the function $f(z)$ defined in 
(\ref{parametrization}) satisfies the differential
equation:
$$[ z(z-1) {d^2 f \over d z^2}
+ (2z-1) {d f \over dz} + {1\over 4} f]=
$$
\begin{equation}
\label{hypergeoapp}
={-{\rm Const.} \over 8} (z-2)z^{-3/2}(z-1)^{-1/2}
\end{equation}
Moreover, the first KZ  equation yields directly
the desired relationship 
(\ref{Fthreeg}).
For ${\rm Const.}=0$ the solutions of (\ref{hypergeoapp})
are the functions recorded in
(\ref{twistconformalblock}).



\begin{references}

\bibitem[*]{permaddress} Permanent address.

\bibitem{Ludwigunpub} A.W.W. Ludwig (unpublished).

\bibitem{Bernard} D. Bernard, A. LeClair, cond-mat/0003075.

\bibitem{MCW} C. Mudry, C. Chamon, X.-G. Wen, Nucl. Phys. B466 (1966) 383.

\bibitem{Tsvelik} J.-S. Caux, N. Taniguchi, A. M. Tsvelik,
Nucl.Phys. B525 (1998) 671-696; (and references therein).

\bibitem{SNR} M. Scheunert, W. Nahm, R. Rittenberg,
J. Math. Phys 18 (1977) 155.

\bibitem{MS}  Z. Maassarani, D. Serban, Nucl. Phys. B489 (1997) 603.

\bibitem{FMS} D. Friedan, E. Martinec, S. Shenker, Nucl. Phys.
B271 (1986) 93.

\bibitem{Gurarie93}
V. Gurarie, Nucl. Phys. B410 (1993) 535.

\bibitem{logs} For recent work on $c=-2$, see e.g. the following,
and referenes therein:
 H. G. Kausch, hep-th/0003029; M. R. Gaberdiel, H.G. Kausch, Phys. Lett.
B 386 (1996) 131;  M. A. I. Flohr, Int. J. Mod. Phys A 12 (1997) 1943.


\bibitem{Kausch}
H.G. Kausch, Nucl. Phys. B583 (2000)513;
and hep-th/9510149.

\bibitem{GLR} I. Gruzberg, A.W.W. Ludwig, N. Read, Phys. Rev. Let. 82 (1999), 4524.

\bibitem{Gurarieb} V. Gurarie,  Nucl. Phys. B546 (1999) 765.


\bibitem{Gradshteyn} For example: I. S. 
Gradshteyn, I. M. Ryzhik, {\it Table of Integrals, Series and Products}
(Academic Press 1980, New York).

\bibitem{GurarieLudwig} V. Gurarie, A.W.W. Ludwig,
cond-mat/9911392.




\end{references}
\end{document}